\documentclass[journal]{IEEEtran}

\usepackage{cite}
\usepackage[utf8]{inputenc}
\usepackage{url}

\usepackage{etoolbox}
\apptocmd{\sloppy}{\hbadness 10000\relax}{}{}

\ifCLASSINFOpdf
  \usepackage[pdftex]{graphicx}
  \graphicspath{{./graphics/}}
  \DeclareGraphicsExtensions{.pdf,.jpeg,.png}
\fi

\ifCLASSOPTIONcompsoc
 \usepackage[caption=false,font=normalsize,labelfont=sf,textfont=sf]{subfig}
\else
 \usepackage[caption=false,font=footnotesize]{subfig}
\fi

\begin{document}

% Just like Nintendo before us and a bunch of other companies and governments,
% let's set abbreviated codenames for this paper!
% MethodicAlly Defeating NintEndo Switch Security or MADNESS.
% While we're at it breaking into the Secure Witch, sounds good to me, right?

% As a side note we wanted to add a mitigation part for each part of the paper but this
% kind of felt off, so we removed them last minute. You can check our drafts in
% iffalse
\title{Methodically Defeating Nintendo Switch Security}

\author{Gauvain Tanguy Henri Gabriel Isidore~Roussel-Tarbouriech,
        Noel~Menard,
        Tyler~True,
        TiniVi,
        Reisyukaku\vspace{-2.0em}}

\maketitle

\begin{abstract}
We explain, step by step, how we strategically circumvented the Nintendo Switch's
system security, from basic userland code execution, to undermining and exposing
the secrets of the security co-processor. 

To this end, we've identified and utilized two distinct analysis procedures.
The software-based analysis suffices for reverse-engineering the userland and
operating system services, and is necessary for a general architectural 
understanding of the software systems in the Nintendo Switch. While this method is
extremely powerful and provides significant leverage over the control of the
system and its software security, a hardware-based method was devised, which
employs analysis of the trusted bootstrap code in ROM. This strategy was essential for
the goal of defeating the hardware root of trust.

Together, these two vectors provide essential insight required to instance a chain
of attacks, in order to gain ROP code execution from the context of a high-security mode
of a secure co-processor of a running system, thus allowing us to demonstrate a
multi-faceted approach on attacking secure, embedded devices in an unfamiliar 
and novel environment.
\end{abstract}


\section{Introduction}

The security of home entertainment devices, such as video game consoles, has been a
critical focus in consumer products engineering and development, as a prerequisite to
enforcing the control and flow of media and advertisement revenue, as well as
protecting end-user data, including personally identifiable and payment information,
and protecting intellectual property on a consumer device. The use of
Digital Rights Management (DRM) techniques, involving cryptographic signatures,
depends on securing digital secrets and title content, and mitigating
against attacks, both hardware and software, local and remote, in an evolving
landscape of platform security challenges.

Video game consoles, in particular, pose a multi-faceted platform security
engineering challenge, with many critical parts maintaining highly-demanding
cryptographical systems, while not compromising the performance of the running
game program. These challenges have made video game consoles a particularly
interesting case study on platform security, through the lens of research.

While older consoles had minimal security, usually to prevent counterfeit and
unlicensed hardware and software from entering the market, more modern consoles
have begun to employ intensified system security methods, usually focused on
protecting the game media from being copied. Although earlier attempts at securing
these consoles have been met with trivial exploits \cite{wii} \cite{ps3}, newer
consoles have employed more serious defenses, and researchers have had to likewise
apply their creative ingenuity, and sometimes discover novel methods \cite{vita}
to find security flaws in these systems.

\subsection*{The Switch, briefly}

The Nintendo Switch is Nintendo's take on state-of-the-art platform security for
embedded devices. It does have all you would expect: a full address-space layout
randomization (ASLR) scheme, hardware-enforced no-execute (NX) to facilitate a
write-xor-execute (W\^{}X) memory policy, sandboxed userland applications,
and a microkernel with a modular, minimally-privileged, cryptographically-marshalled
services architecture enabling strict isolation of services and enforcing a principle
of least privilege. In addition to these software mechanisms, the hardware platform
provides ARM TrustZone capabilities, and a somewhat useful security co-processor,
which are used by a secure monitor program and secure firmware to verify that the
boot path had not been tampered with, and decrypt protected boot programs, applications,
and content.

In March 2017, at the time of the Nintendo Switch release, we began our research,
in order to understand this new device. The first thing that we noticed right away,
is that Nintendo definitely seems to have learned from its past mistakes.

\section{Userland exploitation}

\subsection{Web of (dis)trust}

An interesting point is in how Nintendo dealt with the possibility of browsers being
a target, as they had been in the past, \cite{3ds_browser} \cite{wiiu} by removing
access to the browser altogether, no longer touting it as a feature.

While this is radical, this would have been indeed a somewhat useful mitigation...
if we really had no way to actually launch it. As a matter of fact the browser was
still present, as an applet program for use by the system and games, to provide web-
based or HTML5 content, such as software manuals or tailored online experiences.
However, this browser applet could only connect, through a secure protocol such as
HTTPS, to websites to whom it would have been whitelisted for access. It was also
capable of accessing a limited set of local files. So, this browser seemed rather
useless to our research effort. That is, until you realize that update 2.0.0,
an update available at the release of the Nintendo Switch,
added a notable exception to the restrictions, for authenticating to public Wi-Fi. 

As a handheld console, this makes sense. Allowing the user to connect to public
Wi-Fi access points when, say, waiting for a train, is a function that should
be supported by a mobile device like this. The issue here, is that by only obscuring
the fully-featured web browser behind the captive-portal landing page function,
which is necessary to allow users to authenticate on many public Wi-Fi networks, 
entirely destroys and buries Nintendo's one-pronged approach on browser security
thus far. These Wi-Fi networks often depend on a web browser.

Such schemes, most of the time, do not use any kind of secure 
protocol for the HTTP connection and, even if it did, it would be impractical if not
impossible, to actually go through and whitelist, one by one, all the public Wi-Fi
landing pages in the world. This empowers us to perform a simple server-side
redirection to a web page of choice, one hosted locally and under our control,
and begin to study the system.

\subsection{detachSecurity()}

Web browsers have long been a target for security researchers, as many embedded
devices use web features, and most of those use WebKit for their browser engine, a program
mostly licensed under LGPL. This means that any commercial use of the WebKit engine must
redistribute the source code of the software, including any modifications made to it.
Not only does this allow us to reproduce an instrumentable test environment for debugging
outside of the system, but by its popularity and sheer pervasiveness in the desktop world,
we can easily and quickly know which existing, public and available security vulnerabilities
haven't been patched for our target version of WebKit, and adapt/reimplement them as needed for
this system.

A thing to note, however, is that the WebKit maintainers (Apple) don't just hand us a neat,
preformatted list of security bugs for us to study, as security issues are considered
protected reports. They are removed from public view by the bug tracker, as necessary to allow
for mitigations to develop internally. These reports permanently remain obscured by the bug tracker.
Although, as those bug reports are mentioned in the commits which fix them, we can simply use the
old trick \cite{wiiu} of making a list of all commits linking to bug reports which we do not have
access to, as those are presumably security-related issues of particular interest to us.

On our side, we settled for a use-after-free vulnerability in the FrameTree unload handlers
\cite{detach_vuln} which we named detachSecurity(), as we could simply have a frame DOM object,
"attached" to a detached tree. We then worked on getting some instrumentation in place, by setting
up an RPC server. This let us interact with our exploit remotely to, say, dump current memory.
Soon after, we realized one important thing: the Switch does not have JIT in its browser.

Since the user is not intended to be able to use a web browser at all, the lack of JIT
surprisingly makes sense here. Traditionally, to exploit WebKit, the typical procedure
involved creating some dummy function or functions, running it enough times so that it is
selected to be optimized, wait for it to be compiled and transferred into the JIT memory,
and then, since JIT memory is writable and executable by design, use a vulnerability that
gives you a write in program memory, in order to overwrite said function with our own code,
and execute it. The lack of JIT, and the subsequent lack of executable-and-writeable memory,
completely prevented us from using this technique to gain arbitrary code execution. We had
to resort to the use of Return Oriented Programming (ROP), by way of overwriting function
pointers. While ROP is very powerful in its own right, by letting us redirect program control
flow, it does not let us redirect it to our own code. This primitive level of control,
however, is enough for further research that could lead to privilege escalation, so we went
and began to implement some ROP gadgets --- essentially groups of pointers to functions and
existing executable code fragments, based on our memory dumps, which we can use to effect
changes on the state of the system and test for more powerful vulnerabilities.

\subsection{PegaSwitch}
That period of research was, unfortunately for us, short-lived.
Almost at the same time, an entire toolkit for browser-based exploitation on the Nintendo
Switch was released by another independent team of researchers. The toolkit was
called "PegaSwitch". \cite{pega}

While this was frustrating, as most of our effort up to this point had now seemed worthless,
this was also an invaluable research tool for us, as it implemented all the core ROP functions
we cared about. Whereas we had planned on discovering an applicable WebKit browser
vulnerability, with the expectation that we would have to engineer an exploit stack (including
ROP, privilege escalation, arbitrary code execution, sandbox escape, and shellcode to launch
outside the browser sandbox), PegaSwitch used an already available exploit, namely CVE-2016-4657
\cite{jbme_cve}, commonly known as "Pegasus", with the added twist that this exploit stack was
already implemented for iOS systems on devices with the AArch64 architecture \cite{jbme},
explaining why they were able to get to ROP so quickly and reverse engineer so many things in so
little time, as the Nintendo Switch is also an AArch64 machine.

Now that we have remote userland ROP code execution, our next goal was then to reverse-engineer
the inner workings of the operating system, to identify potential targets for further escalation.
Horizon is the name of the Nintendo Switch's operating system, the kernel, and its system services. 
Horizon is also the OS on the Nintendo 3DS; the Switch version is a further development of this OS,
with refinements and further progress towards being a full-fledged operating system framework.
The 3DS had been thoroughly researched and documented by the time of our research, and
understanding that platform was a great help in understanding the rational of its successor, as
will be explained later.

To the effort of reverse-engineering the OS, we scanned the addressable memory for executable
formats in order to gain useful insight on the fundamental details of the software systems.
One of the first things we realized, apart from the executables being in an apparently new,
custom format --- not uncommon for Nintendo \cite{3ds_formats} --- is that due to dynamic library
linking and loading, we were able to retrieve a bunch of symbols which identified available
services.

\subsection{Services, explained}

What follows is in-depth information about the system service architecture of Nintendo's Horizon
operating system. 
This section in particular is informative; it contains information that we had
to figure out after-the-fact, but is essential to understand the rationale behind our method,
moving forward.

The userland API of the Nintendo Switch is comprised of services, provided by system service
processes, or servers. These are processes which run in the background, and provide
controlled-access, higher-privileged system functions to userland applications as well as other
servers, by way of an IPC interface.

Most of these services are limited as to what they can affect, and their servers load as a
userland program would, and thus have restricted access to other services and hardware, as any
application.

Some of these services are especially privileged, and their servers load very early in the boot
sequence. Because they are loaded so early, before the mechanisms are in place to securely load
and verify executable content from storage, the first six of these servers are contained within
the kernel binary, and load into memory along with the kernel. These kernel-initiated processes,
or KIPs, are compressed and packaged with a unique executable container format, and unpacked by
the kernel once it is loaded and running. Once spawned, these six servers are sufficient for
continuing the secure boot process.

Access to services from userland applications, is coordinated by a service manager (sm),
which is itself a server, with service endpoints. The sm reads the list of allowed services for the
application, registered to sm at launch. This list is in the executable's metadata on storage,
which is signed by Nintendo during their production-signing step, the same for all applications,
games, game cartridges, and system programs for the Switch. That list's signature is verified by the
filesystem server (fs), along with the executable code, to ensure that it has not been tampered
with after installation.

\subsection{IPC and services API, explained}

Nintendo had chosen to use a marshalled, limited code surface in the serving of IPC in the kernel,
and to segregate software risk and performance domains into separate services. Access to each
service endpoint, both by applications and lesser-privileged services, requires different privilege
and access flags, enforced cryptographically by a code-signing trust root.

This is done to further limit the attack surface of kernel-level and privileged code, while not
compromising on the performance of the game program at the forefront. Its kernel is a microkernel,
as stated previously. Through discovery of symbol tables in a dump of early firmware, we found
that the kernel and servers are linked against the same internal library, possibly to minimize code
variances in common security-critical functions, but mostly so that they could follow one known-good
implementation of the rather wild IPC handler code, which we'll explain in detail further on.

To summarize this a little bit better, let us imagine a process wants to use bsd sockets, which
are available as a service under the `bsdsockets` server process. First the process is going to
connect to a port named "sm:", short for Service Manager. It is going to use the IPC command 0,
"Initialize", at this port, to bring up the service frontend for the application, inform the service
of your service access, and allow sm to
ensure that the service is loaded and ready to accept commands, and inform the application that it
is ready. On a successful response from the service manager, the Initialize command is then followed
by IPC command 1, "GetService", with the name of the service we want as an argument. Let's assume
for our example, the service is "bsd:u". We now have a handle to our service and are able to send
commands via that handle, through IPC. A simpler comparison could be made that we interface with the
system and its services through an address-and-port system, in comparison to TCP/IP: the service
name defines where the command is to be addressed; meanwhile the handle, like a TCP port, tells us
where to send a command. The IPC header defines the command format.

\begin{figure}[ht]
  \includegraphics[width=\columnwidth]{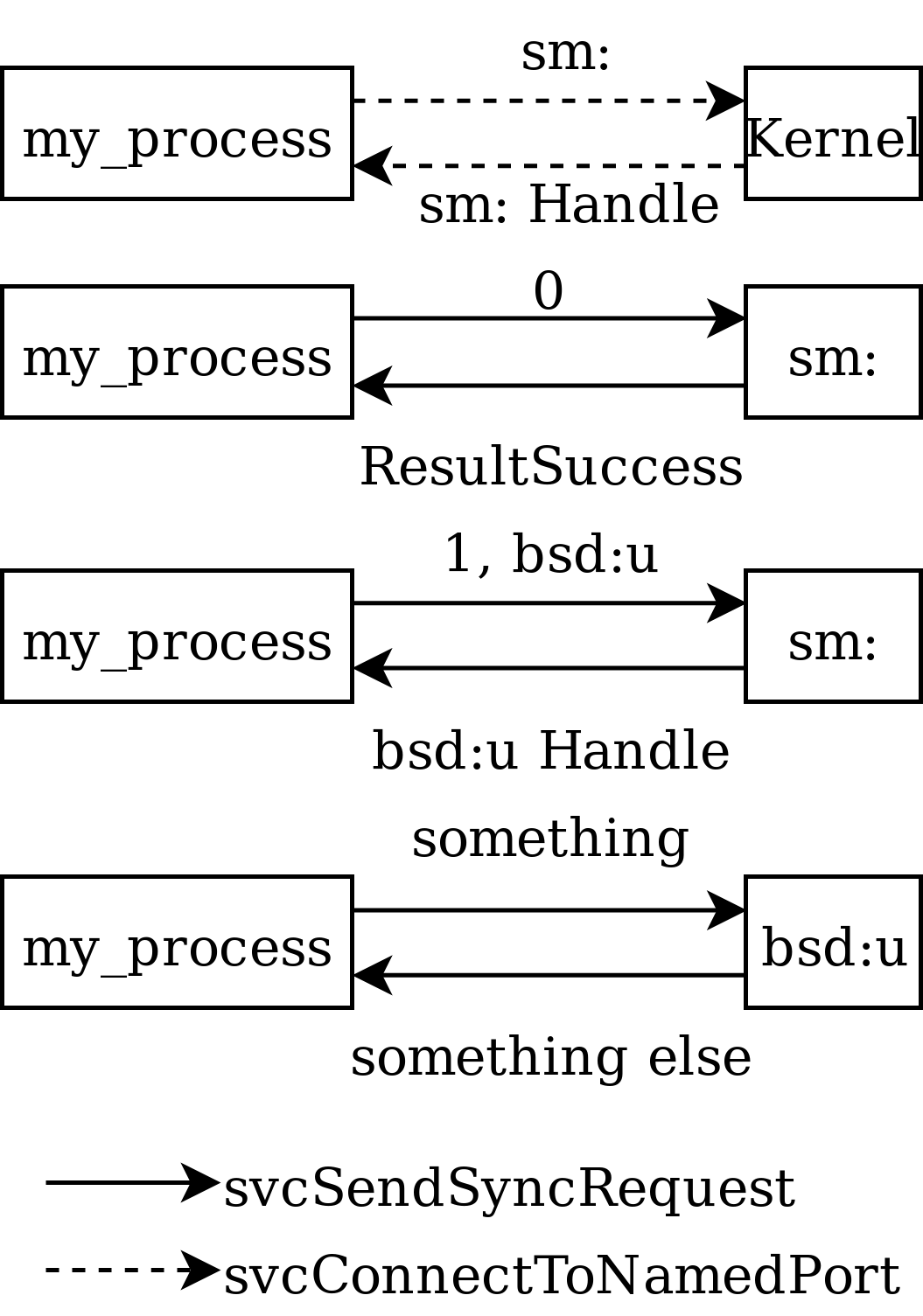}
    \caption{The example above illustrated}
\end{figure}

\subsection{Breaking through: Privileged service access}

As the astute reader may have noticed, not only are service names segmented, e.g., "name:function",
allowing for more fine-tuned isolation, but they are managed by a central service-managing managed
services server, the Service Manager server and its service manager services, which we will get
back to in a moment. The first step in our research was, logically, to understand how to
communicate to those services, in the hope to escalate privileges and reverse-engineer more of
the system. Luckily for us, the architecture hasn't changed much since the days of the
Nintendo 3DS and, having WebKit OSS source code, we could reasonably translate the unincorporated 
SDK library calls to service and IPC patterns with the help of the 3DS and its webkit binaries,
which include the data necessary. Through this effort, we were able to identify and reverse engineer
a bunch of service related functions. However, unfortunately, most of the services' names, functions
and the IPC command structure seemed to have been changed entirely, making us do the documentation
effort and tooling all over again.

Reversing the port system was trivial, as it was a simple syscall away --- namely, poking at SVC
0x1f, "svcConnectToNamedPort"\cite{syscalls}, was enough to verify that the named ports enumerated
prior were correct, and to enable us to further manipulate the services behind these ports.
Gaining a meaningful understanding of the IPC command structure was, to put it mildly, a fair
bit more painful. By "a fair bit", of course, we regret to inform you, it had been head bangingly
stupidly obfuscated, and blanketed in seemingly nonsensical optimizations.
The more dedicated (read: masochistic) reader may wish to consult the additional documentation
about the IPC buffer descriptors, and the bit shuffling hacks that happen behind the scenes, to
completely grasp the magnitude of the topic at hand \cite{ipc_madness}.

Let us assume we somehow reverse engineered this protocol and reimplemented it.
As we do not have any way to dump privileged code yet, our first reflex was to
fuzz the services to find any evicence of a crash thanks to an unexpected input, while documenting
them... or I should say, it should have been our first option. Remember how I mentioned the Service
Manager up above? If you follow the rules, you are supposed to call an Initialize function, which
sets the PID of the current process to handle service permissions. It turns out, you can just
skip the initialize function.
Doing that, leaves the PID field uninitialized. Uninitialized fields are set to 0. PID
below 8 have unrestricted service access, which we could also deduce at this point since
the 3DS worked in a similar fashion \cite{pre_smh}.
It should suffice that the name given to this exploit, "sm:h", need not be further explained.

\subsection*{In brief:}
\begin{itemize}
\item Features that are able to be abused, will be abused, even if they're not "supposed" to be.
\item Completely removing such a feature may not be the easiest solution, and it may not be the
best for all use-cases, but it is the most secure.
\item Sometimes compromises are necessary to meet demands, but be sure to investigate all
alternatives before settling with the lesser-secure option.
\item Security through obscurity is not secure. Sometimes the very fact that something is hidden,
is more than enough information to formulate a plan of action. (This goes to both the Wi-Fi browser
access, and the WebKit bug reports.)
\item Just because something can't make code executable, doesn't mean it can't execute code. ROP
is a very powerful exploitation technique.
\item If you reuse common libraries, code and programs from other platforms, you run the risk of
importing existing vulnerabilities and exploits from those platforms, too.
\item If the prior system is documented, its successor will be easier to figure out. This can be
used to the benefit of both protecting, and exploiting, the system.
\item Stripping executables of their debug strings and symbols makes the adversary's job harder.
\item A comprehensive audit strategy on privileged code's API endpoints is crucial.
\end{itemize}

\section{Privilege Escalation: System Services}

\subsection{Following a lead}

All system service processes except for the kernel-initiated servers, are contained in "sysmodules"
in storage, like the 3DS.

By this point we had a huge attack surface, nothing short of unrestricted
access to all system services, which we could take advantage of to find flaws in privileged
processes. However, there remained much mystery behind the workings of the system. We had not
dumped the sysmodules yet, and thus could not perform out-of-system analysis on these binaries.

We began automating the process of fuzzing some services, and then kind of got disinterested from
research temporarily. In the meanwhile both Switchbrew and Reswitched independently found an exploit
called "pl:utonium". This exploit was in the first commands of the service pl:u (which handles the
system shared font in shared memory), initialized by the ns server, which read from an array using
the user-input arguments. That input was obviously not sanitized prior to use, allowing them to
dump the entire binary of the NS sysmodule \cite{3ds_flaws}. ReSwitched, on their side, created an
emulator to automate the findings of vulnerabilities through fuzzing \cite{ctu}
\cite{mephisto}. Right after this, we had a surge of renewed interest into the
Switch which made us investigate some of the highest privileged sysmodules,
as they would be the most useful to break.

%%% Removed because of politics
\iffalse
Before telling you how we decided to look at entrypoints, I would like to
mention that we independently confirmed that at least one research group was able to be aware
of privileged services through internal leaks they shouldn't have had access to,
greatly helping their ability to document and research said modules.
\fi

Moving on, we were aware that Plutoo was able to dump system modules \cite{modules_dump}
while excluding the possibility of a kernel hack \cite{kernel_modules}. This led
us to believe that the flaw was present in some of the more privileged system
servers, namely the ones built-in that Plutoo mentioned. We were also curious
on why plutoo listed both the title id and the name of the modules in the post\cite{modules_dump}
mentioning the dump, and, while this could have been a coincidence, we decided to look at what
this could lead us to.

As such we looked at the list of title ids that were currently documented on
switchbrew \cite{switch_title_list} and enumerated all the privileged ones: fs, ldr,
ncm, pm, sm and finally, boot. 
As the sm server's services have a very small attack surface with very few commands, and we had
already bypassed it by way of sm:h, we decided to not look at it. 
The boot server being a "headless" server without any kind of service frontend, exploiting it
was out of the question. 

\subsection{The FS services}

After removing those two from the equation, we decided to look at everything
filesystem related, because of the peculiar name listing in the post that plutoo made
\cite{modules_dump}. And so we began trying to work on fs, as it was
numerically the first built-in, fully-privileged module that was also related to filesystems.

We began looking at the FS documentation, and studying and exploring every possible entrypoint in
the service set handled by the FS server. Coincidentally, the first service port listed was
fsp-ldr, along with its first command ("OpenCodeFileSystem", though it was referred to as
"MountCode" back then, due to a lack of debugging symbols for naming internal things).

Unfortunately, trying to bind to it directly, throws an error. We had somewhat anticipated this;
building upon our experience with, and knowledge of, the 3DS, we figured that there was the
likelihood that this service had a session limit, and that said limit was occupied at
initialization time, something which happened to the FS counterpart of the 3DS
\cite{3ds_single_session}. As the service name was fsp-ldr (which we presume stands for "privileged
filesystem service pertaining to the loader"), we figured out that if we crashed the ldr service,
which, one could infer, had an exclusive handle to fsp-ldr... we could get access to the fsp-ldr
handle instead!

And as a matter of fact, this is what happened: Any method of crashing, killing, stopping,
unloading, or otherwise causing a denial-of-service attack to ldr (of which there are several),
would cause the ldr server to release its handle on fsp-ldr, which we could hook up to, and then
ask fsp-ldr nicely to dump all the code modules it had access to, since by virtue of its function,
it needs access to the binaries for applications and sysmodules alike. Soon after we figured this
out, a description of the vulnerability, a minimal proof-of-concept exploit, and a functional ldr
DoS were released to the public by other researchers \cite{switch}. We found out that they had left
out enough details so as to not completely trivialize the exploit. As we found, there was no
reason to keep the full-featured exploit private, and hoarding that exploit would only hurt the
further-developing public research effort, so we ported our unabridged fsp-ldr exploit code to the
DoS framework that they had released and published it, including it into PegaSwitch
\cite{dump_modules_gov}.

\subsection*{In brief:}
\begin{itemize}
\item While separating services from more privileged code in the same program is a noble attempt
at preventing leakage, and marshalling the service domain and its endpoint away from the core
program helps... again, a comprehensive audit strategy can prevent faults in privileged code.
\item Just as parallellizing computation speeds results, so does parallellizing research.
\item While dependency chains are often hell on system designers, asserting a dependency such that,
for example, fsp-ldr can only operate when ldr is present, would have kept fsp-ldr from being
exploited. 
\end{itemize}

\section{Beyond Privilege Escalation: Boot and Trust}

\subsection{Approaching the Tegra}

The following section is written under the assumption that we never did any work on the Tegra X1
hardware prior to beginning the research effort with the Nintendo Switch. We had, however, already
been working on the Tegra X1 platform, since around August 2016.
Unfortunately, we cannot discuss this prior research, as it is held under NDA until further notice.

We believe that such antagonistic NDA terms needlessly restrict innovation and hamper
security for the company in question, as well as harming the professional security research
community and vendors who implement Tegra-based solutions. We will say that at least one major
vendor that makes use of Tegra X1-based platforms, abuses these non-disclosure terms to the
detriment of their researchers.

Having a way to dump almost all of the system modules, we began to look at our options to
escalate privileges in a more concrete, reproducible and persistent fashion. Having the binaries
had helped a ton, and while we did get somewhere with our newfound information, we were stopped
dead in our tracks by another research team, fail0verflow, who had shown off a cold-boot exploit
\cite{oh_come_on_it_was_getting_fun}, hinting at the fact that they were able to dump the bootrom.

At this point we were aware that a development kit for the Tegra X1, the System-on-Chip (SoC) used
by the Nintendo Switch, was publicly available for purchase... and that it most likely had the same
bootrom. Thus, the fun part began: glitching and hardware fault engineering was put into play.
We will mostly skip over this part, as Yifan Lu already explained very well how glitching could be
applied to embedded devices in the past \cite{vita} \cite{vita_glitch} and an entire talk had
already been given at the C4 OpenChaos event on glitching the Tegra X1, specifically.
\cite{switch_glitch}.

Essentially, we took the path of least resistance, and voltage-glitched a Tegra X1 development kit,
and a second device which we unfortunately cannot discuss here due to the aforementioned NDA.
As we are unable to discuss that project, we will assume that our research is entirely unique
and that fail0verflow inspired us to look into the bootrom, for the purposes of this paper.
We will try to release the research currently under NDA as soon as possible. We are sorry for any
mandatory omissions we do in this paper, and are hopeful that the available literature will be more
than sufficient to satiate the curiosity of the dedicated reader.

\subsection{Analyzing the boot ROM}

So, we glitched our device and successfully acquired the contents of the boot IROM. At this point,
since we were aware of fail0verflow's tweet \cite{oh_come_on_it_was_getting_fun}, we saw that they
had, purposefully or not, hid the USB port of the console. Understanding that the USB protocol ---
especially USB 3 and USB Type-C --- has a level of implementation complexity that often stretches
beyond the definition of "acceptable", and on the heels of several USB kernel flaws released a bit
before this tweet \cite{usb_very_bad}, we surmised that it was fairly likely that the flaw was
related to USB in some way. At this point we had only gained a fairly minimal understanding of the
Tegra boot firmware logic, and as such, decided to write an emulator for it, and employed dynamic
program analysis using that emulator, to aid the static analysis effort.

Building this emulator not only forced us to understand the boot flow of the Tegra X1, including
its recovery mode (RCM) which makes use of USB, but it also made us understand way better how the
boot ROM code worked. It had helped so much, that we actually found the bug in a minimal amount of
time, with a minimal amount of effort, with our emulator still only operating in HLE
(High Level Emulation)!. 

The plan we had in mind for this emulator was to, first of all, get an idea of the bootflow by
implementing all the main cryptographic/USB functions in HLE, in order to develop a tool that would
interface with the console's RCM. We already knew that, besides the RCM interface, there wasn't a
way to interface to the console's boot ROM to provide input, external to the console. Once we had
a better definition of the RCM interface over USB, we would then make a basic fuzzing tool that
would run in the background over a long time, while we ported our emulator to full Low Level
Emulation (LLE), to allow us to more completely simulate the processor and its devices, to
thoroughly fuzz out any trivial bugs we may have missed up to this point.

While we were working on the HLE part of our emulator, we decided to perform a quick audit.
The USB and cryptographic functions, being prime targets, were the first items of interest to us,
with some focus on cryptographic fails. Fortunately for us, the bug occured at a point in the RCM
program flow before any kind of cryptographic verification was performed. As such we discovered it
naturally when trying to understand the RCM.

\begin{figure}[ht]
  \includegraphics[width=\columnwidth]{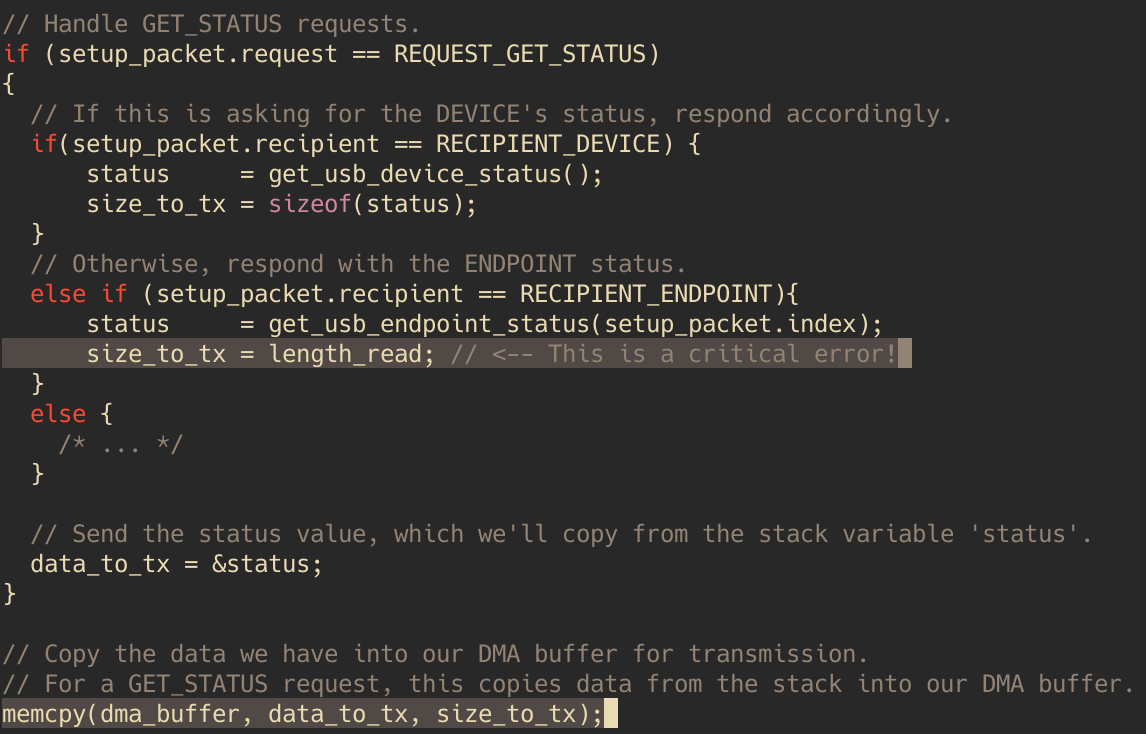}
    \caption{The bug, courtesy of ktemkin's fusée gelée report \cite{fusee_gelee}}
\end{figure}

The flaw was fairly simple actually: in some functions of the USB protocol, we could arbitrarily
control the size of a memcpy, allowing us a good, simple buffer overflow, just like the 90s, with
no mitigation whatsoever, no memory protection or isolation, as this is a bootrom, naturally, with
no operating system or kernel to enforce it with.
The memcpy occurs before the stack, so you just overwrite the stack pointer, point it to the code
you copied, and since there are no stack cookies, no ASLR, and absent no-execute memory policy,
you just watch it execute your code. It's so easy, it almost feels like cheating.

\subsection{God's in his TrustZone, all's right with the world}

Having access to an exploit in boot ROM, we had no real incentive to work on the switch again, but
we were curious about some known but private vulnerabilities nonetheless, especially one that the
ReSwitched group called "déjà vu". To put that into context: ReSwitched, early on, publicized a
write-up about a flaw they found called "jamais vu", which allowed code execution on the Secure
Monitor of the Nintendo Switch \cite{jamais_vu}, while announcing déjà vu. As this is the type of
exploit that really makes use of all the unexplored intricacies of the system, we will more
thoroughly explain the overall technical architecture of the Nintendo Switch.

The Nintendo Switch is composed, first and foremost, of a SoC called the Tegra X1, created by
Nvidia. While it may sound unintuitive, this SoC is actually composed of several processors with
different architectures and different use-cases, out of which some are particularly notable.

The main CPU core complex, that we will henceforth call the CCPLEX, is the primary applications
processor. The "Boot and Power Management Processor", referred to in this paper as BPMP or
BPMP-Lite, handles system bringup, power management, and is technically the "root processor" of a
Tegra system. (The Reference Manual calls it "BPMP-Lite", as it lacks some features that more
advanced versions of the SoC will apparently get.) The boot ROM that we dumped before, is referred
to by Nvidia as the BPMP-FW, the "firmware" for this subsystem, because it is the first program
loaded and the first processor to initialize on power-up.
There is also a third "core complex", called the Tegra Security 
Co-Processor (TSEC) powered by a Falcon microprocessor, that we will talk more in depth about in the following section. 
For the purpose of this section, knowing that the BPMP is
meant to handle power-on bringup, initial bootloader, and low-level tasks is more than enough.

The CCPLEX is a somewhat recent ARM processor, which means it is capable of a feature that ARM
calls TrustZone. For those unfamiliar with the concept, TrustZone is a hardware-enforced virtualized
system-separated enclave on the processor, used to isolate security-critical parts of the operating
system as much as possible (in the case of the Nintendo Switch, its internal cryptographic engine).
This introduces a notion of "Secure World" and "Normal World", both running their own OS and having
their own separate resources. For example, the Secure World has its own secure RAM space, "TZRAM".

In our case, the "Normal World" is the Horizon kernel, with its servers and userland applications,
and the "Secure World" contains the "Secure Monitor" of the Nintendo Switch, which is just its
cryptographic engine, as mentioned above, alongside some rudimentary power management services.
The normal world interacts with this secure world by using Secure Monitor Calls (SMC), roughly
analogous to kernel syscalls, or "Supervisor Calls" (SVC). This is an important part of Nintendo's
security scheme, as this allows them to seal keys, even in the case of complete kernel takeover,
so that:
\begin{enumerate}
   \item We cannot replicate their cryptographic engine outside of the device, and
   \item They can always patch known vulnerabilities, update the keys and we would
have no way to break the DRM of newer games.
\end{enumerate}

\begin{figure}[ht]
  \includegraphics[width=\columnwidth]{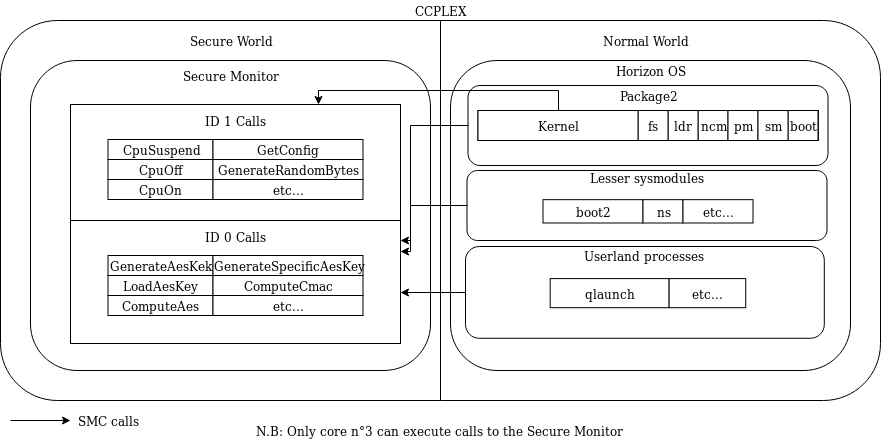}
    \caption{TrustZone illustration}
\end{figure}

Unfortunately for them, having bootrom code execution kind of spoils the fun... but let's just
forget that for a moment. Continuing on what we mentioned above, we know that ReSwitched managed
to get code execution on the Secure Monitor which they called "déjà vu", on top of "jamais vu",
which pointed us in the right direction, especially thanks to the amazing write-up for the latter.

We were still pretty dormant up until firmware revision 6.0.0 of the Nintendo Switch.
The founder of, and a former lead of the ReSwitched group had independently discovered and
privately disclosed a vulnerability to Nintendo that was part of the exploit chain that formed
déjà vu \cite{nvhax}, and as of 6.0, Nintendo had also mitigated the exploit itself! \cite{6.0}
This was more than enough to bring back interest to this bug, even though we would have no real
use for it.

\subsection{Waking up in another world}

So, after all this fuss, what is Jamais Vu all about? Essentially, it deals with deep sleep --- a
mode which "shuts down" all hardware, except for the minimal power draw necessary for sustaining the
Power Management Controller (PMC), and persisting the contents of DRAM. When resuming from deep
sleep, the BPMP will begin as if being cold-booted, but it instead resumes the system state in RAM.

As mentioned in the excellent write-up cited above \cite{jamais_vu}, in earlier firmwares,
before entering deep sleep, the Secure Monitor would save its state from its TZRAM, to the CCPLEX's
"main" DRAM. It saves the contents encrypted, along with a Message authentication code (MAC) to
ensure that it had not been modified or tampered with. The MAC is saved to a set of hardware
registers in the Power Management Controller (PMC), which persist through deep-sleep. The idea is
that, when rebooted, unless you had bootrom code execution, you would not be able to modify the
Secure Monitor whatsoever, keeping it secure and encrypted, so that no information leaks
from Secure World can happen.

There was, unfortunately for Nintendo, a small omission in that trust logic: userland could modify
the PMC registers, breaking entirely the "trust" in TrustZone. We could simply replace the copy
of the TZRAM with our own, change the MAC result to verify against ours and, thanks to another
bug within the boot ROM, because it fails to correctly verify the state of the Security Engine (SE)
it restores, we can simply replace the keys used to decrypt the TZRAM by ones we control, to
make it decrypt and verify our own TZRAM.

On top of all of that, we could have modified BPMP exception vectors to point at code we owned,
leading to pre-sleep code execution on the BPMP.
We could have rewritten the RESET vector to execute our code at startup and, as such, be able to
control the Secure World like before. This didn't last long, though, as all of this would be fixed
with a launch-day firmware update, 2.0.0.

Or at least this is what Nintendo thought: Indeed, while PMC was effectively made Secure-Mode only
as far as we can tell, the BPMP exception vectors checks haven't been thoroughly reviewed:
a high speed internal bus of the Tegra X1, the AHB, has a DMA interface, conveniently named AHB-DMA.
AHB-DMA is supposedly "deprecated", but it still exists in the X1. Because this DMA hadn't been
otherwise disabled or limited to specific memory ranges, we could overwrite the exception vectors
of the BPMP, leading to a full compromise once again.
The good news, is that the AHB-DMA interface is only accessible to the kernel, running on CCPLEX.
The bad news, is that it is available to the CCPLEX, let alone non-Secure World kernel code, in the
first place.

\subsection*{In brief:}
\begin{itemize}
\item Glitch attacks, fault injection, power analysis, and other hardware-level attacks, will
violate your preconceptions of software/hardware validity at the lowest level, unless you
proactively protect against unexpected code or register data injection.
\item Complexity (such as, full implementations of bus and network stacks, RSA cryptosystems,
complex implementations of storage interfaces, filesystems and other device driver code) increases
the potential vulnerability surface. Paradoxically, the more complex the program, the easier it
becomes to exploit, in general.
\item Low-level boot-time resources are a very high focus for researchers to audit. Not only the
bootloader, but recovery modes, factory modes, download modes, whatever they may be called, if it's
loading code into the system from an external source, it is going to be thoroughly studied for any
possible flaws.
\item Moving data between security domains (TZRAM to DRAM, for instance) is another high-focus area
to audit. Anything that depends on securely saving and restoring such data should not depend on an
unsecured system (PMC registers, for instance) to protect them.
\item "Undocumented" features aren't... or won't be for long, in the context of security research.
\item Further, "Deprecated" means nothing to a security researcher, until the "deprecated" feature
is removed.
\item Validating the state of privileged boot services goes a long way to enforcing secure boot.
Payload signatures aren't enough; further boot-time measurements of code and state are necessary.
\item Stating that an exploit exists is as good as publishing it; if you don't, someone else will.
\end{itemize}

\section{Beyond Trust: from TSEC to \texttt{0xDEAD5EC1}}

Having bootrom-level code execution, we thought we were pretty much done with the security
challenges of this console... And yet, Nintendo still managed to surprise us.

Before we touch on how Nintendo revamped their trust chain, we'll take a look at some background on
console security.

\subsection{How Homebrew Works}

The community of developers and users making and using unlicensed homebrew software for game
consoles, depends on the ability to run arbitrary, unauthorized code in a convenient way (e.g.,
without physically modifying the device or installing aftermarket or replacement components through
high-risk hardware modifications). In order to run unauthorized code, the strict code-signature
enforcement on applications needs to be relaxed, if not removed entirely. On older consoles and
handheld systems, there was no code-signature enforcement; running homebrew applications depended
entirely on gaining any arbitrary code execution, most commonly through specially-crafted
user-generated data or savedata. On later consoles, cryptographical measures had taken place to
simultaneously enable game data to be stored on removable, user-accessible memory, as well as to
protect the game data from tampering, reverse-engineering, and outright replacement, and prevent
unauthorized applications from being able to run on users' systems. Needless to say, these measures
are often defeated by security researchers, and made available to homebrew developers and users.
In past systems with code-signature enforcement, these workarounds had been achieved through 
gaining code execution, and once running on the target, installing modified or patched versions of
the firmware or operating system, typically called "custom firmware", or CFW. For instance,
installing FlashMe on the Nintendo DS allows running unsigned code directly over the Download Play
feature, through a protocol termed "Wireless Multiboot".

However, the ability to modify the system firmware or OS contents, had given rise to a handful of
malware programs, often disguised as pirated games or highly anticipated homebrew applications.
Trojan/DSBrick.A, one such piece of malware, simply displays a brick-wall texture on the displays
of the DS system, while it overwrites the system firmware, rendering the console entirely useless
(a "brick"), requiring risky or costly repairs to return the system to a usable state.
As well, earlier Nintendo Wii homebrew methods involved replacing IOS packages (essentially, system
drivers) with custom ones enabling homebrew to have access to more hardware that only few games
had taken advantage of. These "cIOS", custom IOS packages, were high-risk modifications, and often
led to bricked Wii systems, especially if modified IOSes were installed before accepting a system
update from Nintendo.

Because of the risk of these and other modifications to the system potentially leading to damage,
customer support headaches, and in the case of DSBrick, media attention, console manufacturers
had to make a compromise to further secure against system modifications, while still allowing
upgrades to firmware, system software, bundled applications, downloaded applications and content.

Because of the heightened protections on firmware modifications, homebrew methods on later consoles
have given rise to live, in-memory patching, rather than modifying the necessary code in storage,
either through modified bootloaders (such as Enso \cite{enso}, for the PlayStation Vita) or through custom
kernel modules or system programs which patch memory as well as add features
(Prometheus/Pro-CFW \cite{pro_psp},
for the PlayStation Portable), or a combination of the two (Luma3DS \cite{luma}, for the Nintendo 3DS).
By tradition, these pre-loaders, in-memory patchers, modified or reimplemented components
continue to be called "CFW", despite often being neither custom, nor firmware.
These CFW environments almost universally disable or work around code-signature enforcement.

\subsection{Switching gears}

Earlier on, Nintendo Switch homebrew was rudimentary. The limited homebrew entrypoints available
had placed many restrictions on the capabilities of such software, when compared to native, signed
applications. The first public homebrew user and development environment, ReSwitched's PegaSwitch,
which ran on system software 3.0.0, was used to launch the Homebrew Menu, or hbmenu, an alternative
launcher, loader and host for homebrew applications compiled as relocatable code objects
(.nro files). This menu was intended to be bootstrapped from an existing application that had been
exploited to run arbitrary code, which either had, or had escalated to obtain, the necessary
permissions to read content from storage, and dynamically load executable code.

The application from which homebrew was bootstrapped, in this early environment, was the
WiFiWebAuthApplet, or the captive-portal landing page web browser applet. As such, being an applet,
the programs were extremely restricted with the amount of memory they could allocate, as applets
ran in the foreground, while an application (such as a video game) was either running or suspended,
in the background. As well, the web browser had only very minimal permissions, and while this
limited environment would more than suffice for testing new toolchains and enabling homebrew
development on the platform, the Switch is a very powerful system and users were looking to develop
more advanced homebrew, such as PC game engines, emulators, and even creative tools, which would
take advantage of the hightened permissions and larger memory allocation a full application context
can provide.

In order to run more advanced homebrew applications, a CFW environment for Horizon became a
requisite, since such an environment would be able to further enable homebrew access to the system.
The currently state-of-the-art implementation, Atmosphère \cite{atmosphere}, actually
reimplements several system modules and servers, the secure monitor, the bootloader, and parts of
the kernel-initiated processes, and plans to reimplement the entire kernel of Horizon later down
the road, making it an actual CFW.

The important thing is that we are within the environment of Nintendo's firmware, and as such,
we depend on their cryptosystem. As Stratosphère, Atmosphère's sysmodule reimplementation, is not
entirely complete. It currently patches the boot code package, Package2; and as such, needs to be
able to decrypt it.
Even if, by some miracle, the ongoing community effort eventually allowed us to entirely
reimplement all of Horizon, we would still hit a wall when it comes to actually playing Switch
games, as we would need a complete reimplementation of their cryptosystem. While
this is doable, the 6.2.0 update introduced new security features barring us
from accessing secret hardware keys, necessary to reimplement the cryptosystem, 
as explained more in detail below. 

Thus, we need to be able to derive keys on our own to be able to maintain the current features
of this console. And this changed in the 6.2.0 update. To understand why this is important,
we are going to explain the Nintendo Switch boot flow and cryptosystem in more detail below.

\subsection{Ignition, Switch}

Beginning the boot flow, the BPMP of the SoC powers up, launching its bootrom. This bootrom, 
depending on the state of PMC registers, either performs a warmboot and loads from existing state
in DRAM, or loads and verifies package1, and jumps to package1ldr. In that case, package1ldr takes
care of decrypting and verifying PK11. Package1, containing package1ldr and PK11 are stored in the
first eMMC boot partition, and PK11 contains the warmboot binary, NX-Bootloader, and the secure
monitor firmware. The warmboot binary is what is saved to DRAM by the secure monitor when entering
the deep-sleep state.

After package1ldr does its initialization, it then jumps to the bootloader within PK11, called
NX-Bootloader, which in turn loads and launches the Horizon OS main kernel and modules. When
the boot server process has initialized, it triggers a command in the process manager (pm), 
causing it to load and launch the boot2 server, which is the first non-built-in system server.
For the sake of simplicity, it is shown in the figure below that boot launches boot2 directly.
boot2 then takes care of launching all the system servers; one of which, Nintendo Shell (ns),
launches the main front-facing user menu, qlaunch.

\begin{figure}[ht]
  \includegraphics[width=\columnwidth]{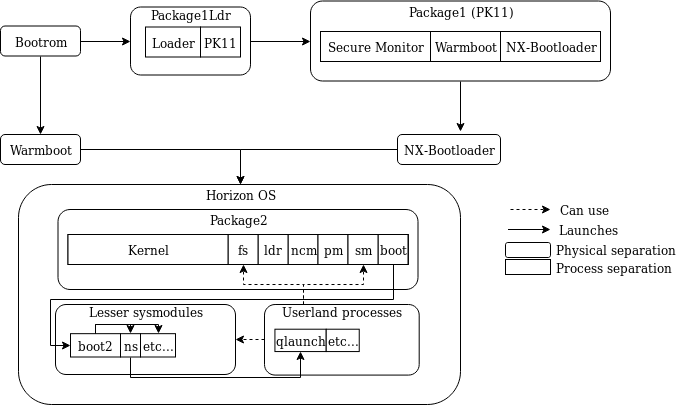}
    \caption{Nintendo Switch boot flow}
\end{figure}

At the lowest level, prior to the 6.2.0 firmware update, the entire cryptosystem of the Nintendo
Switch relied on 2 keys: The SecureBootKey (SBK)) and TSEC key. The first
key is set by the boot ROM and is console unique, while the second one was generated from
hardware secrets on the security co-processor, still console unique, which, coupled with 
more console uniques keys on the eMMC boot partitions, generates static keys used for 
everything after this. The TSEC firmware was loaded by NX-Bootloader, and we could, at this 
point, simply run the firmware blob and read back our result from 
the SOR1 registers, which were used as a secure transfer route
between the TSEC and bootloader. We could not reimplement this firmware either, as only
authenticated code, and thus signed by an approved authority, would have access to the hardware
secrets on the TSEC. This firmware was of course largely ignored up until the 6.2.0 update.

\subsection{The "S" in Switch is for "Secure"}

The 6.2.0 update introduced major changes to the way key generation worked, and most
importantly, the TSEC firmwares. Up to this point the firmware had 3 main stages,
Boot, KeygenLdr and Keygen. As mentioned earlier, TSEC firmwares can be signed,
but they can also be encrypted, and as such KeygenLdr decrypted Keygen using
hardware secrets we did not have access to, as KeygenLdr was signed.
The update introduced two new stages to this whole chain, SecureBootLdr and
SecureBoot. The entire TSEC boot chain had thus been reconstructed to the following flow:
Boot $\rightarrow$ SecureBootLdr $\rightarrow$ KeygenLdr $\rightarrow$ Keygen $\rightarrow$
SecureBoot. 

While KeygenLdr and Keygen haven't been updated, SecureBoot added some interesting security
concepts. Not only does it generate yet another key from hardware secrets, the TSEC root key,
but this time it tries to prevent us from actually getting the contents of the SecureBoot binary
through simple means, such as by halting the BPMP, rewriting its exception vectors, and dropping
the BPMP instruction pointer back into code it controls (a signed and encrypted Package1).
This is actually fairly interesting! We cannot replace any TSEC firmware
blobs with any of our code, because we are unable to sign it, and we can't remove it because that
code would generate the new required keys from secrets we don't have... This forces us to redirect
program flow to code trusted by Nintendo, making this an adept attempt to re-secure its console,
even after a critical boot firmware bug.

At least, this would be the case, if we weren't able to simply fool TSEC into assuming everything is
fine, and that we didn't just take control from it.
We can do just that, because otherwise what good would a security co-processor be, right?
We can use the BPMP's control over the internal I/O memory management unit, the System Memory
Management Unit (SMMU) as the ARM architecture calls it, to redirect all reads and writes to pages
of memory we control, and fool it into thinking it effectively redirected code flow from our own
control, while it just handed over the keys. That sure is a foolproof way to go about in-depth
platform security.

But Nintendo was not done with TSEC, or at least not yet. As it turns out, the TSEC had a feature
that was not well-documented (some would argue that it isn't documented at all, for our usecase),
that came in handy for securing the boot flow and giving us
yet another new challenge. An SMMU Bypass function is available to the TSEC, which forces the TSEC
to simply consider all memory as linear, and avoid the memory virtualization that SMMU can perform.
That's quite a useful feature for a security coprocessor, we must admit. This time around, in
firmware update 7.0.0, they enabled it, while doing a bunch of checks to detect virtualization,
updating yet again their TSEC root key so as not to let us use older ones.

\subsection{In-"sept"-tion}

Since 7.0.0 had yet again changed the security playing field, we needed to find another approach
to attack TSEC.  To explain our findings, we will discuss how this firmware authenticates its
payloads in depth.

TSEC is based around a Falcon microcontroller. This processor has three "modes" in a security
context: Non-Secure (NS), which typically restricts the microprogram from reading most registers
and memory; Light Secure (LS), which is rarely used outside of debugging and development; and
Heavy Secure (HS), which enables full access to the cryptographical hardware, and protected or
secret registers and memory.

A small blob of unauthenticated data is present in the firmware uploaded to the
TSEC. that data contains the size of each payload, and the AES-CMAC (!) that should be
calculated for the payload to pass boot-safety measurements.

Now, if you understand even elementary cryptographical theory, that seems counter-intuitive. Why
AES-CMAC, and not an asymmetric cypher like RSA? If you get the AES key used for decryption (and
the one necessary for verification), you can effectively sign your own MAC... and completely substitute the
expected payload for one you have control of! Well, Nintendo has a reason. Falcon is a rather
limited microcontroller environment. For one, there's very few memory headroom to (securely)
implement RSA in software. Falcon has separate data and instruction memory, DMEM and IMEM, and
together they measure in the dozens of kilobytes. For another, one would have to implement RSA
from scratch to the Falcon microarchitecture, and Nintendo seems to have since learned not to roll
their own crypto... However, the TSEC does include hardware AES acceleration, which is expected
to be reasonably secure and is a convenient way to get firmware-specific keys
following the Falcon specification.

Those checks also execute in reverse order, so that KeygenLdr can ensure that Boot has not
been tampered with, for example. On top of this CMAC, TSEC verifies page by page that any
signed payload the payload is indeed signed, at the hardware level, before granting it the
HS privileges. Those pages are then marked secret and cannot be read anymore
until CPU halt, where trying to read from it would return \texttt{0xDEAD5EC1}. 

Now that we've covered that, let's assume we are working with earlier TSEC firmwares, for brevity.
Those are easier to work with, while maintaining, thanks to the nature of our exploit, the exact
same level of control.

Keygenldr, during normal operation, reads that blob of data, so that it can parse it and check the
first stage size and CMAC hash, in our case Boot. This is done in order to retroactively verify
that it hadn't been tampered with. The thing is, this blob of data is not authenticated.
We can control both the size and content of data being copied over, hence, we can control a
rudimentary stack smash. To go about that, we can have KeygenLdr copy our modified boot payload,
fail verification, and return. Fortunately for us the verified MAC has not been
cleared from memory, so we can restart the process along with this MAC, pass
verification and return to our crafted ROP gadgets, allowing us to get ROP code execution in the
Heavy Secure mode of the TSEC.

The exploit described above is used to get ROP code execution after Keygen has been
decrypted, but its pages have been marked secret. We could optionally get it 
after the verification has failed, but we would be unable to decrypt Keygen. We
would also like to note that the same result could have been replicated thanks
to design flaws of the TSEC that we shall not further discuss here.

\subsection*{In brief:}
\begin{itemize}
\item The community behind developing homebrew software is pervasive; methods to gain homebrew
access are often invasive, and often co-opted by software pirates.
\item It is very difficult to secure a system where the boot chain of trust has been compromised.
\item Adding a new cryptographically-secure system may seem like a good protection against such a
compromise moving forward, but keep in mind that added complexity makes for a larger attack surface.
\item Understand that using symmetric algorithms (such as AES) where asymmetric crypto would
typically be employed, means that a compromise of the key or keystream will allow forging the data
in question.
\item If you're depending on unauthenticated data from within a high-security domain, even if it
is only accessible by highly-privileged code, exercise due
diligence on ensuring that the data is valid --- e.g., employ bounds-checking on structures in the
data.
\end{itemize}

As a side note, we would like to warn the interested reader that, should they decide to further
attempt to understand the secure boot process of the Switch by reading the publicly available
Atmosphère \cite{atmosphere} reference code, take note that the key-generation is not done for
7.0.0 firmwares in the same way that it is done on the original bootloader.

Sept \cite{sept}, is a payload they designed to bypass checks implemented in SecureBoot. It does so
by loading the original 7.0.0 TSEC firmware, unmodified. The TSEC firmware is designed to verify
the AES CMAC of the PK11 binary, before returning execution to the bootloader there. 
Sept works because the CMAC was forged on the custom PK11, passing the code authentication routines
in SecureBoot. Sept, now in PK11, derives keys in place in package1Ldr
and scrambles the TSEC and TSEC root key, making it impossible whatsoever to use the original
Secure Monitor firmware, as it doesn't have the keys necessary to further perform key generation.

\section{Conclusion}
We completely broke the security system of one of the most secure embedded 
consumer devices on the market with no prior knowledge of its hardware nor
software.   

Unlike most of the existing literature on computer science security, we decided to focus on the 
inductive process of finding security flaws and fixing them, rather than explaining how they work
and implementing exploits for them. This decision was motivated by having most of the security
flaws we independently found already published online, some being released before or even
during \cite{8.0.0} the writing of this paper! Our inability to talk about some potentially
damaging non-public security vulnerabilities related to the Tegra X1 did not help either.
We also think finding flaws in embedded devices is much more interesting to write about,
as a whole. 

We would also like to stress that this paper's ultimate goal is not to expose Nintendo's flaws,
but rather to help computer security research be aware of such possible flaws, and as such would
like to give our point of view on what could have been improved by Nintendo to avoid those exploits.

Firstly, exploits such as sm:h and pl:utonium seem to present a crucial lack of
in depth auditing.
While we are aware that auditing the entire runtime code of the Switch firmware
would be impractical, serious security audits must be done on every privileged
bit of code that could be harmful and are a prime attack target, which is indeed the 
case of the NS and SM servers. Optionally, switching to safer languages (such as Rust
\cite{rust}) or formally-verified coding paradigms, would have altogether avoided both of those
flaws.  If some code has to be privileged and yet not trusted, such as a blob of third
party code, then it is good practice to try to isolate it as much as possible. Positioning
services such as pl:u separately from the main NS services is also a good idea, even
though such separation should have been more pronounced in that particular case.

Moving on to any secure code attacks and secure key retrieval: it is crucial to add anti glitch
measures to make glitching with low-cost, low-complexity equipment as hard as possible.
Depending on your threat model, it might also be a good idea to encrypt the ROM stored inside the
SoC. While a powerful attacker could, ultimately, decap the SoC, reverse engineer
cryptography primitives used to decrypt the boot ROM \cite{tv_hack}, and laser glitch his way
around the program, the process should be as arduous as possible if we want any cryptographically
secure software boot on devices at all. Decap and laser glitching will most likely always be
possible, but when this becomes part of the least effort approach then we might have some chance at
protecting secrets.

Any code implementing any cryptography primitive, especially if it's considered secure and crucial
to the chain of trust in a system, should not only be audited in order to avoid basic security
issues, but also be hardened against side-channel attacks, such as power analysis and fault
injection \cite{yifan_dfa}, at the very least, to avoid any early breakage of the trust chain.

Ultimately, security is defined by failures thereof. Software Engineering should learn from
traditional engineering in that point, as software and software systems CAN fail, and we should
accordingly plan for even a slim margin of error in that regard. As such, mitigations against
common faults in software security should be graciously applied whenever possible, and a clear list
of everything a user could potentially control, must be set. 

A sad thing about security, is that it is mostly decided by your budget.
Even if you have a state-level intelligence threat the smaller your budget is, the less you'll
invest in security, if this isn't one of your unique selling points. 
This shows in IoT devices and, in some cases, forced hackers to exploit vulnerabilities, in order to
patch them. Our approach to security should change altogether.

We must focus on putting the best practices at the forefront, to the point that they are the easiest
to implement, otherwise we will continue to be trapped in the conspiracy against trust.

\section*{Acknowledgments}
We would like to thank the tremendous work of the research groups that worked
on the security and making of the homebrew community on the Nintendo Switch:
Switchbrew ReSwitched and, to a lesser extent, fail0verflow. We acknowledge
the time and effort investment to make such a community exist. While we were
only interested on the security side of the subject, the entire SDKs, CFW
reimplementations and homebrews are invaluable for such a community. The entire
documentation provided by Switchbrew contributors was also immensely helpful at
several points during this research. We would also like to thank 3DBrew, which
community is shared by Switchbrew, who were a building block for Switchbrew
research.

\bibliography{madness}
\bibliographystyle{IEEEtran}

% i think do_crypto(cf switchbrew) is a nice function, don't you?
\end{document}